\documentclass{article}
\usepackage{graphicx}
\usepackage{url}

\title{Computational Geometry Column 37}
\author{%
  Erik D. Demaine%
    \thanks{Dept.\ of Computer Science, Univ. of Waterloo,
      Waterloo, Ontario N2L 3G1, Canada.
      \texttt{eddemaine@\penalty \exhyphenpenalty uwaterloo.ca}.}
\and
  Joseph O'Rourke%
    \thanks{Dept.\ of Computer Science, Smith College, Northampton, MA
      01063, USA.  \texttt{orourke@cs.smith.edu}.
      Supported by NSF Grant CCR-9731804.}}
\date{}

\let\latexcite=\cite
\def\cite{\nolinebreak\latexcite}
\let\latexref=\ref
\def\ref{\nolinebreak\latexref}

\usepackage{amsfonts}
\newcommand\R{\mathbb{R}}

\begin{document}
\maketitle
\pagestyle{empty}
\thispagestyle{empty}

\begin{abstract}
Open problems from the 15th Annual ACM Symposium on Computational Geometry.
\end{abstract}

Following is a list of the problems presented 
on June 15, 1999
at the open-problem session
of the 15th Symposium on Computational Geometry.

\begin{description}
\item[Jack Snoeyink, Univ. North Carolina, snoeyink@cs.ubc.ca]~

  Are there $n$ points in $\R^3$ that have more than $n!$ different
  triangulated spheres?

  An upper bound of $O(10^n n!)$ derives from Tutte's bounds on unlabeled planar
  triangulations times the ways to embed vertices, ignoring intersections.
  A lower bound of $\Omega( (n/3)! )$ is established by a construction with $n/3$
  thin needles.  

    Frequently in computer graphics, triangulated spheres are represented
    through two separate components: geometry (vertex coordinates) and
    topology (vertex connectivity).
    A negative answer to the posed question permits the compression
    of the topological information by storing it implicitly within the geometric
    information.

\item[Jeff Erickson, Univ. Illinois at Urbana-Champaign,
      jeffe@cs.uiuc.edu]~

  Say that a nonconvex polyhedron in $\R^3$ is \emph{smooth} if (a)~every
  facet is a triangle with constant aspect ratio
(i.e., the triangles are {\em fat}), and (b)~the minimum dihedral
  angle (either internal or external) is a constant.  Can any smooth polyhedron
  be triangulated 
  (perhaps employing Steiner points)
  using only $O(n)$ tetrahedra?  Equivalently, is it always
  possible to decompose a smooth polyhedron into $O(n)$ convex pieces?

  Both conditions (a) and (b) are necessary.  Chazelle's polyhedron~[C84],
  which cannot be decomposed into fewer than $\Omega(n^2)$ convex pieces, has
  large dihedral angles, but its facets have aspect ratio $\Omega(n)$.  
  A variant of Chazelle's polyhedron can be constructed with fat triangular facets, but
  with minimum dihedral angle $O(1/n^2)$.

  Smooth polyhedra are not necessarily ``fat''\footnote{
Under any reasonable definition of {\em fat}, e.g., a bounded volume
ratio of the smallest enclosing sphere to the largest enclosed sphere.
}---for example, a $1\times
  n\times n$ rectangular ``pizza box'' can be approximated arbitrarily
  closely by a smooth polyhedron with $\Theta(n)$ facets.  The
  Sch\"onhardt polyhedron (a nontriangulatable twisted triangular prism~[E97])
  is smooth, so smooth polyhedra cannot necessarily be triangulated
  without Steiner points.  In fact, gluing $O(n)$ Sch\"onhardt
  polyhedra onto a sphere yields a smooth polyhedron that satisfies
  both conditions and requires $\Omega(n)$ Steiner points.  Is this the
  worst possible?

  \begin{description}
  \item[{[C84]}] B.~Chazelle.  Convex partitions of polyhedra: A lower bound
    and worst-case optimal algorithm. \emph{SIAM J. Comput.}
    13:488--507, 1984.
  \item[{[E97]}] D.~Eppstein.  Three untetrahedralizable objects.  Part of
        \emph{The Geometry Junkyard}.  May 1997.
        \url{http://www.ics.uci.edu/~eppstein/junkyard/untetra/}
  \end{description}

\item[S\'andor P. Fekete, TU Berlin, fekete@math.tu-berlin.de]~

  What is the complexity of the Maximum TSP for Euclidean
  distances in the plane?

  Barvinok et al.~[BJWW98]
  have shown that the \emph{Maximum TSP} (i.e., the
  problem of finding a traveling salesman tour of maximum length)
  can be solved in polynomial time,
  provided that distances are computed according to a
  polyhedral norm in $\R^d$, for some fixed $d$. The most natural
  instance of this class of problems arises for rectilinear distances
  in the plane $\R^2$, where the unit ball is a square.
  With the help of some additional improvements by Tamir,
  the method by Barvinok et al.\ yields an $O(n^2\log n)$
  algorithm for this case.
  This was improved in [F99] to
  an $O(n)$ algorithm that finds an optimal solution.

  For the case of Euclidean distances
  in $\R^d$ for any fixed $d\geq 3$, it has been shown [F99]
  that the Maximum TSP is NP-hard. The case of $d=2$
  remains open, but the problem poser conjectures it to be NP-hard.

  \begin{description}
  \item[{[BJWW98]}]
    A.~I.~Barvinok, D.~S.~Johnson, G.~J.~Woeginger, and R.~Woodroofe.
    The maximum traveling salesman problem under polyhedral norms.
    {\em Proc. 6th Internat. Integer Program. Combin. Optim. Conf.},
    Springer-Verlag, Lecture Notes in Comput. Sci. 1412, 1998, 195--201.

  \item[{[F99]}]
    S.~P.~Fekete. Simplicity and hardness of the Maximum Traveling Salesman
    Problem under geometric distances.
    {\em Proc. 10th Annu. ACM-SIAM Sympos. Discrete
    Algorithms}, 1999, 337--345.
  \end{description}

\item[Greg Perkins, Rutgers Univ., gperkins@paul.rutgers.edu]~

  The ``wireless communications location problem'' is
  to determine the point $p$ on the $z=0$ plane at which a user is located
  from signals detected by a collection of receivers.
  The user sends out a radio signal (a sphere expanding at uniform velocity), 
  which perhaps diffracts around or reflects off of buildings before
  being received.
  The given data is as follows:
    \begin{itemize}
    \item A set $B$ of orthogonal 3D boxes (the buildings), 
     oriented isothetically,
     each with its base on the $z=0$ plane, each taller than the highest
     receiver;
    \item A set $R$ of receiver locations (points in $\R^3$); and
    \item A set $S$ of received signals, each described
      by the receiver and the distance traveled along the signal's 
      polygonal path.
    \end{itemize}

The signals $S$ come in three varieties:
\begin{enumerate}
\item Line-of-sight (LOS) signals, direct from $p$ to $r \in R$.
\item $1$-diffracted signals, diffracted at most once around an
edge or vertex of some building $b \in B$.  In effect the expanding
sphere of radio waves is retransmitted at building corners.
\item $k$-reflected signals, reflected at most $k$ times
off of building walls (which act as a radio-wave mirrors).  
It may be assumed that $k \leq 3$.
\end{enumerate}

LOS signals can be distinguished from diffracted and reflected signals.
It may be assumed that no signal is both diffracted and reflected.
Various pragmatic assumptions may be made about the maximum propagation distance
of a signal, and the size of the buildings with respect to this distance;
see~[PS99] and~[F96] for the technical details.
An approximate location for $p$ with bounded error would suffice.
A natural question is determining necessary conditions for
accurate location of the user.
Preprocessing can be as extensive as needed.

\begin{description}
\item[{[PS99]}]
G. Perkins and D. Souvaine. 
Efficient radio wave front propagation for
wireless radio signal prediction.
DIMACS Tech. Rep. 99-27, May 1999.\\
\url{http://dimacs.rutgers.edu/TechnicalReports/1999.html}

\item[{[F96]}]
S. Fortune.
A beam-tracing algorithm for prediction of indoor radio propagation.
{\em Proc. 1st ACM Workshop on Appl. Comput. Geom.}
Eds. M. Lin and D. Manocha.
Springer-Verlag, Lecture Notes in Comput. Sci. 1148, 1996, 76--81.
\end{description}

\item[Mark Overmars, Utrecht Univ., markov@cs.uu.nl]~

Define a {\em prism\/} $\mathcal{P}$ as a polyhedron 
formed by extruding a polygon $P$ in the $z=0$ plane to the $z=1$ plane.
Call the bottom and top faces $P_0$ and $P_1$ respectively.
Given two different triangulations of $P_0$ and $P_1$, 
each perhaps using Steiner points,
determine whether $\mathcal{P}$ can be tetrahedralized
respecting the triangulations of $P_0$ and $P_1$,
perhaps with Steiner points, but using only $O(n)$ tetrahedra.
If so, provide an algorithm to find such a tetrahedralization.

\item[Kasturi Varadarajan, Rutgers Univ., krv@dimacs.rutgers.edu]~

  Is there a topological cube with orthogonal opposite facets?  More
  precisely, does there exist a bounded convex polytope in $\R^3$, whose graph
  ($1$-skeleton)
  is the same as that of a unit cube (6 quadrangular facets with
  the incidence relationships of a cube), 
such that for each of the three pairs of
  opposite facets, the planes containing them form a dihedral
  angle of 90 degrees?

  \mbox{%
  \vbox{\hsize=4in
  During the open-problem session, John Conway observed that there is a
  nonconvex topological cube with this property.  His construction is as
  follows.  Take three quadrangles arranged in the plane as shown to the right,
  and add three vertical rectangular faces at the bold edges,
meeting at a vertex at infinity.  Then ``distort''
  this polyhedron slightly to make a topological cube with orthogonal
  opposite faces.  The center vertex dents inwards, so the polyhedron is
  nonconvex.}%
  \vbox{\hsize=1in
  \hspace*{0.29in}\includegraphics[scale=0.8]{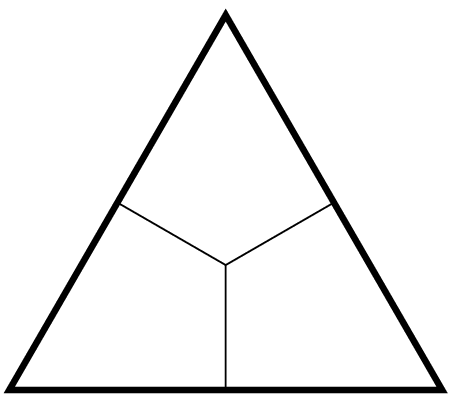}}%
  }

\item[Mark de Berg, Utrecht Univ., markdb@cs.uu.nl]~

  Solve either of the following problems in $o(n^4)$ time:

    \begin{enumerate}
     \item Let $S$ be a set of $n$ segments in the plane.
          Determine whether there is a point $p$ seeing all segments
          completely.
     \item Let $P$ be a set of $n$ points in the plane. Find a point $q$ in the plane
   that maximizes the minimum difference between any pair of distances $d(p,q)$
   and $d(r,q)$ over all pairs $p,r\in P$.
   That is, find a point $q$ that maximizes
   $\min_{p,r\in P} | d(p,q) - d(r,q) |$.

    \end{enumerate}

The goal of the second problem is to reduce the dimension of $P$
from two to one by replacing every point by its distance to the
point $q$.
For this to be effective, $q$ should be chosen so that
it discriminates between any pair of points as much as possible.

\item[Suresh Venkatasubramanian, Stanford Univ., suresh@cs.stanford.edu]~

  Let $P$ be a set of $n$ points in the plane.  Call a unit circle
  \emph{good} if it touches at least three points from $P$.
  What is the maximum number of good unit circles that can be placed,
  maximized over all possible sets of $n$ points?

  All that is known is a trivial upper bound of $O(n^2)$, and a lower bound of
  $2n$ (on a lattice).  
This problem was mentioned in [B83];
nothing more recent is known.
It has an application in estimating the
  complexity of certain pattern-matching problems.

\begin{description}
  \item[{[B83]}]
J. Beck.
On the lattice property of the plane and some problems of
Dirac, Motzkin and Erd\H{o}s in combinatorial geometry. 
{\em Combinatorica} 3(3--4): 281--297, 1983.
\end{description}

\item[John Conway, Princeton Univ., conway@math.princeton.edu]~

  Several classic problems were reposed: 

    \begin{description}
    \item[Angel Problem:]

       Consider the following game played on an infinite chessboard.
       In each move, the \emph{angel} can fly to any uneaten square
       within 1,000 king's moves; and the \emph{devil} can eat any one square
       (while the angel is aloft).
       The angel wins if it has a strategy so that it can always move;
       otherwise the devil wins.
       Prove who wins.
       {\em Reward\/}: \$1,000 if the devil wins; \$100 if the angel wins.

\begin{description}
  \item[{[BCG82]}]
E.~R.~Berlekamp, J.~H.~Conway, R.~K.~Guy.
    {\em Winning Ways for Your Mathematical Plays}.
    Volume 2: {\em Games in Particular}.
    Academic Press, 1982, p.~609.
\end{description}

    \item[Thrackle Problem:]
       \emph{Thrackles} are made of ``spots'' (points in $\R^2$) and
       ``paths'' (smooth closed curves, ending at spots), with the condition
       that any two paths intersect at exactly one point, and have distinct
       tangents at that point.
       Can there be more paths than spots?
       {\em Reward\/}: \$1,000.

\begin{description}
  \item[{[O95]}]
J. O'Rourke.
Computational geometry column 26.
{\em Internat. J. Comput. Geom. Appl.} 5:339--341, 1995.
Also in {\em SIGACT News\/} 26(2):15--17, 1995.
\end{description}


    \item[Holyhedron Problem:]

       Is there a polyhedron with a hole in every face
       (i.e., every face of which is multiply connected)?
       {\em Reward\/}: \$10,000 / number of faces in the discovered polyhedron.
       The reason for the divisor in the reward is that a solution
       is known for an astronomical number of faces.

    \item[Danzer's Problem:]

       Given an infinite set $S$ of points in $\R^2$,
       with the property that there are at most $k$ points of $S$ 
       in each
       disk of radius $1$,
       must there be arbitrarily large convex ``holes,''
       i.e., regions containing no points of $S$?

\begin{description}
\item[{[CFG91]}]
H. P. Croft, K. J. Falconer, and R. K. Guy.
{\em Unsolved Problems in Geometry}.
Springer-Verlag, 1991.  Problem E14.
\end{description}

    \end{description}

\end{description}

\end{document}